\newcommand{\Tr}{\mathop{\rm Tr}\nolimits}
\def\slash#1{\setbox0=\hbox{$#1$}               
   \dimen0=\wd0                                 
   \setbox1=\hbox{/} \dimen1=\wd1               
   \ifdim\dimen0>\dimen1                        
      \rlap{\hbox to \dimen0{\hfil/\hfil}}      
      #1                                        
   \else                                        
      \rlap{\hbox to \dimen1{\hfil$#1$\hfil}}   
      /                                         
   \fi}                                         %

\documentclass[12pt]{article}
\usepackage{epsfig}
\usepackage{cite}
\usepackage{amsmath}

\begin{document}
\begin{titlepage}
\begin{flushright}
UM-TH-96-16\\
October 1996\\
\end{flushright}
\vskip 2cm
\begin{center}
{\large\bf ${\cal O}(N_f\alpha^2)$ Corrections
           in Low-Energy Electroweak Processes}
\vskip 1cm
{\large R. Akhoury,\ \ \ Paresh Malde,\ \ \ Robin G. Stuart}
\vskip 1cm
{\it Randall Physics Laboratory,\\
 University of Michigan,\\
 Ann Arbor, MI 48109-1120,\\
 USA\\}
\end{center}
\vskip .5cm
\begin{abstract}
We provide a compendium of techniques that can be used to compute
${\cal O}(N_f\alpha^2)$ corrections to low-energy
electroweak processes. Specifically, these are the 2-loop electroweak
corrections containing a light fermion loop. It is shown that the vast
majority of such corrections can be reduced
to expressions involving a universal master integral for which an exact
analytic form in dimensional regularization is given. The only exceptions
are certain photon vacuum polarization diagrams. Examples
are presented for diagrams that occur in a variety of processes of practical
interest.
\end{abstract}
\vskip 4cm
\end{titlepage}

\setcounter{footnote}{0}
\setcounter{page}{2}
\setcounter{section}{0}
\newpage

\section{Introduction}

Among the most precisely measured electroweak observables
the electromagnetic coupling constant, $\alpha$,
the muon decay constant, $G_\mu$, the anomalous magnetic moment $g-2$,
and the ratio of charged to neutral current cross-sections
in semi-leptonic neutrino scattering, feature prominently. Because of their
high experimental precision $\alpha$ and $G_\mu$ are commonly used as
input in calculations of electroweak radiative corrections. Until relatively
recently the experimental errors on $\alpha$ and $G_\mu$ were entirely
negligible compared to those on the third input quantity, $M_Z$. Now,
however, both $G_\mu$ and $M_Z$ are both measured to 2 parts in $10^5$.
Any further improvement on the accuracy of $M_Z$ would mean that the
measurement of the muon lifetime in principle limits the accuracy with
which electroweak theory can be tested.

In this arena, theoretical calculations currently lag experiment.
Calculations are not available that could exploit the
$2\times10^{-5}$ accuracy available on $G_\mu$ and $M_Z$. It is therefore
of technical interest and practical necessity to develop electroweak
calculations that approach the $2\times 10^{-5}$ accuracy of experiment.
While a full 2-loop calculation would be ideal, it is likely that the analytic
form of such results would be too complicated to usefully write down.
One would therefore probably wish to rely on computer algebra and
integration to obtain numerical results \cite{Mainz}. However it
is still true that certain dominant subclasses of 2-loop diagrams can be
calculated and written down in compact analytic form
\cite{BijHoogeveen,Barbieri,DennHollLamp,Fleischer}.
Such results are, of course, always preferable to numerical ones.

The set of 2-loop Feynman diagrams containing one fermion loop are such
a class. These we will refer to as ${\cal O}(N_f\alpha^2)$ corrections where
$N_f$ is the number of fermions and $N_g=N_f/8$ is the number of complete
generations. In these corrections we will assume that fermions are massless
which is an excellent approximation for all but the third generation. The
${\cal O}(N_f\alpha^2)$ results can be used either with $N_g=3$ and mass
corrections computed for the third generation or $N_g=2$ and corrections for
the third generation computed separately.

Because $N_f$ is quite large --- up to 24 --- the ${\cal O}(N_f\alpha^2)$
corrections are expected to be a dominant subset of 2-loop graphs and
since $N_f$ provides a unique tag, the complete set of ${\cal O}(N_f\alpha^2)$
corrections contributing to a particular physical process will form a
gauge-invariant set. Diagrams of this type have been considered in connection
with the muon anomalous magnetic moment \cite{CKM1}.

{\it A priori\/} the ${\cal O}(N_f\alpha^2)$ corrections can be expected to
contribute at the level of $1.5\times10^{-4}$. Without their calculation and
inclusion, theoretical predictions are uncertain at this level.

To test the Standard Model, one needs at least four well-measured electroweak
observables. Three of these are used as input to the model and the fourth
is then predicted and can be compared with experimental determinations. In
addition, relating electroweak parameters, such as $M_Z$, extracted at high
energies with those, such as $\alpha$ and $G_\mu$, extracted at low energies
inevitably involves an uncertainty arising from the hadronic contribution to
the photon vacuum polarization, $\Pi_{\gamma\gamma}^\prime(0)$. This hadronic
uncertainty can be reduced by improved measurements of the cross-section
$\sigma(e^+e^-\rightarrow{\rm hadrons})$ and the use of dispersion relations
\cite{Zeppenfeld,Jegerlehner,Swartz} or eliminated completely by
sacrificing one high precision observable \cite{StuartHadron}.

At present the fourth well-measured electroweak observable, needed to test the
Standard Model at very high precision, is missing but it is not unlikely that
quantities, such as the polarization asymmetry, $A_{LR}$, or $g-2$ of the
muon, could fulfill this r\^ole in the foreseeable future. In any event the
calculation of the ${\cal O}(N_f\alpha^2)$ electroweak radiative corrections
are a necessary prerequisite to be able exploit the extremely accurate
experimental measurement that are the legacy of LEP.

\section{The Master Integral}

In what follows the general coupling of a fermion to a vector boson will
be denoted
\[
\raisebox{-0.7cm}{\epsfig{file=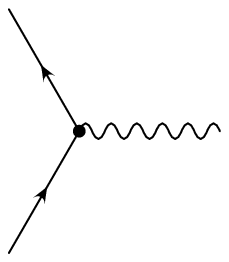,width=1.5cm}}
\equiv i\gamma_\mu(\beta_L\gamma_L+\beta_R\gamma_R)
\]
where $\gamma_L$ and $\gamma_R$ are the usual left- and right-handed
helicity projection operators and $\beta_L$ and $\beta_R$ are the
corresponding coupling constants. We assume throughout an anticommuting
$\gamma_5$. A Euclidean metric with the square of time-like momenta being
negative will be used and all calculations will be done in $R_{\xi=1}$ gauge.
The sine and cosine of the weak mixing angle
$\theta_W$ will be denoted $s_\theta$ and $c_\theta$ respectively.

We will be interested in the ${\cal O}(N_f\alpha^2)$ corrections for massless
fermions. In corresponding Feynman diagrams the fermions couple only to
vector bosons and never to Goldstones or the physical Higgs particle.
Many, but by no means all, such diagrams are obtained simply by inserting
a fermion loop into the boson propagator of a one-loop diagram,
see Fig.1a. Because the original one-loop diagram is logarithmically
divergent, the fermion loop insertion will need to be calculated to
${\cal O}(n-4)$ in dimensional regularization, where $n$ is the dimension
of space-time. It turns out, however, that for all ${\cal O}(N_f\alpha^2)$
diagrams for low-energy processes, it is possible to obtain expressions that
are exact in $n$.

For the massless fermion loop insertion, it may be shown that
\begin{eqnarray}
\raisebox{-0.3cm}{\epsfig{file=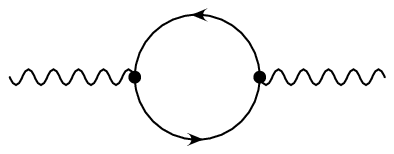,width=2.5cm}}
&=&-\left(\delta_{\mu\nu}-\frac{p_\mu p_\nu}{p^2}\right)
   (\beta_L\beta_L^\prime+\beta_R\beta_R^\prime)
   \frac{\displaystyle (n-2)}{\displaystyle (n-1)}
   \int\frac{d^n q}{\displaystyle i\pi^2}\frac{p^2}{q^2(q+p)^2}\\
&=&-\left(p^2\delta_{\mu\nu}-p_\mu p_\nu\right)
   (\beta_L\beta_L^\prime+\beta_R\beta_R^\prime)
   4(\pi p^2)^{\frac{n}{2}-2}
             \frac{\Gamma^2\!\left(\frac{\displaystyle n}
                                        {\displaystyle 2}\right)}
                  {\Gamma(n)}
             \Gamma\left(2-\frac{\displaystyle n}
                                {\displaystyle 2}\right)
\end{eqnarray}
where $\beta_L$, $\beta_R$ and $\beta_L^\prime$, $\beta_R^\prime$ of the
couplings of the attached vector bosons.

A number of special cases of low energy 2-loop integrals have appeared in
the literature \cite{BijVeltman,FleischerTarasov,DavyTausk}.
By using the projection operator techniques described in the following
sections, most ${\cal O}(N_f\alpha^2)$ diagrams occurring in
low-energy electroweak processes, including box diagrams,
can be reduced to expressions involving a general
master integral of the form
\begin{equation}
I(j,k,l,m,n,M^2)=
\int\frac{d^np}{i\pi^2}\frac{1}{[p^2]^j[p^2+M^2]^k}
\int\frac{d^nq}{i\pi^2}\frac{1}{[q^2]^l[(q+p)^2]^m}
\label{eq:MasterIntDef}
\end{equation}
provided the diagram contains at least one massive boson.
The only exceptions to this are certain contributions to the photon
vacuum polarization. In that case some diagrams must be calculated retaining
the fermion mass, $m_f$, and then applying the asymptotic expansion of
ref.\cite{DavyTausk} before taking the limit $m_f\rightarrow0$.
Other ${\cal O}(N_f\alpha^2)$ diagrams that cannot be
expressed in terms of the integral (\ref{eq:MasterIntDef}) are of the
pure QED type containing only internal photons and
fermions. The relevant master integrals for this case can be found in
ref.\cite{FleischerTarasov}.

In eq.(\ref{eq:MasterIntDef}), the integration over $q$ can be performed
using standard Feynman parameter techniques and yields
\begin{equation}
\int\frac{d^nq}{i\pi^2}\frac{1}{[q^2]^l[(q+p)^2]^m}=
\frac{\pi^{\frac{n}{2}-2}}{[p^2]^{l+m-\frac{n}{2}}}
\frac{\Gamma\left(l+m-\frac{\textstyle n}{\textstyle 2}\right)
      \Gamma\left(\frac{\textstyle n}{\textstyle 2}-l\right)
      \Gamma\left(\frac{\textstyle n}{\textstyle 2}-m\right)}
     {\Gamma(l)\Gamma(m)\Gamma(n-l-m)}.
\end{equation}
The resulting integral with respect to $p$ in eq.(\ref{eq:MasterIntDef})
is independent of angle and hence
\begin{align}
\int\frac{d^np}{i\pi^2}\ \frac{1}{[p^2]^{j+l+m-\frac{n}{2}}[p^2+M^2]^k}
=\frac{2\pi^{\frac{n}{2}-2}}
        {\Gamma\left(\frac{\textstyle n}{\textstyle 2}\right)}
  &\int_0^\infty dp\ \frac{p^{2n-2j-2l-2m-1}}{[p^2+M^2]^k}\nonumber\\
=\frac{\pi^{\frac{n}{2}-2}}{(M^2)^{k+j+l+m-n}}
  &\frac{\Gamma(n-j-l-m)\Gamma(k+j+l+m-n)}
        {\Gamma\left(\frac{\textstyle n}{\textstyle 2}\right)\Gamma(k)}
\nonumber\\
\end{align}
from which it follows
\begin{eqnarray}
 & &I(j,k,l,m,n,M^2)=
    \frac{\pi^{n-4}}{(M^2)^{k+j+l+m-n}}\nonumber\\
 & &\ \ \ \times
    \frac{\Gamma(n-j-l-m)\Gamma(k+j+l+m-n)
          \Gamma\left(l+m-\frac{\textstyle n}{\textstyle 2}\right)
          \Gamma\left(\frac{\textstyle n}{\textstyle 2}-l\right)
          \Gamma\left(\frac{\textstyle n}{\textstyle 2}-m\right)}
         {\Gamma\left(\frac{\textstyle n}{\textstyle 2}\right)
          \Gamma(k)\Gamma(l)\Gamma(m)\Gamma(n-l-m)}
\nonumber\\
\label{eq:MasterIntexpr}
\end{eqnarray}

\section{Self-energy Diagrams}

For electroweak processes, such as muon decay, in which the external
fermions can be considered massless, only the self-energy diagrams of vector
bosons need to be considered. General techniques are available to reduce the
tensor integrals that occur to standard form factors
\cite{PassarinoVeltman}
for which scalar integral representations are known. For the present case
of zero external momentum, $q=0$, the vector boson self-energies,
$\Pi_{\mu\nu}(q^2)$, can only take the form
\[
\Pi_{\mu\nu}(0)=\delta_{\mu\nu}F
\]
where $F$ is a function of the internal masses only and may be obtained
from the tensor integral representation of $\Pi_{\mu\nu}(0)$ by means of
the projection operator, $\delta_{\mu\nu}/n$. Thus
\begin{equation}
F=\left(\frac{\delta_{\mu\nu}}{n}\right)\Pi_{\mu\nu}(0).
\end{equation}
The resulting scalar integral can always written in terms the master integral,
$I(j,k,l,m,n,M^2)$, of eq.(\ref{eq:MasterIntexpr}) and results obtained that
are exact for all $n$.

\section{Vertex Diagrams}

Again for processes with massless external fermions the only relevant
vertex corrections are those involving vector bosons and these will
necessarily be purely vector and axial-vector in character. A general vertex
correction can then be represented as
\[
\raisebox{-0.7cm}{\epsfig{file=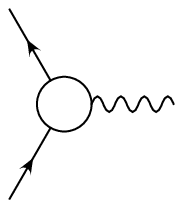,width=1.5cm}}
\equiv V_\mu=i\gamma_\mu(V_L\gamma_L+V_R\gamma_R)
\]
where $V_L$ and $V_R$ are functions only of the internal masses. The tensor
integral representation of $V_\mu$ can easily be obtained by standard
techniques and from it the scalar integral representations of $V_L$ and
$V_R$ follow by means of projection operators. Thus
\begin{equation}
V_{L,R}=-\frac{i}{2n}\Tr\{V_\mu\gamma_\mu\gamma_{R,L}\}.
\label{eq:vertexproj}
\end{equation}
where $\Tr\{\gamma_\mu\gamma_\mu\}=4n$ is assumed.
This method for directly obtaining the scalar integral representation
of the vertex form factors is particularly convenient when computer
algebra is being employed. Once again the resulting scalar integrals can
be written in terms of the master integral, $I(j,k,l,m,n,M^2)$.

\subsection{Example}

\begin{figure}[t]
\epsfig{file=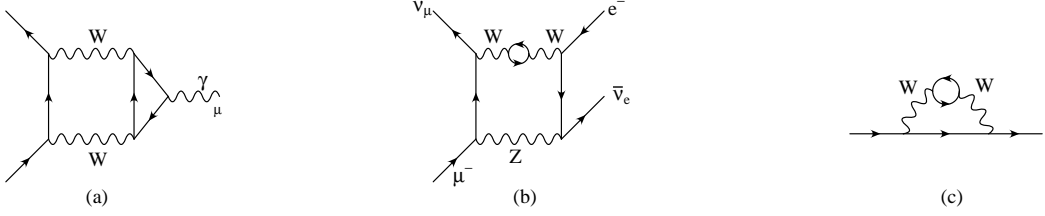,height=3cm}
\caption{Feynman diagrams treated in the examples.}
\label{fig:example}
\end{figure}

The diagram shown in Fig.\ref{fig:example}(a) occurs in the calculation of the
${\cal O}(N_f\alpha^2)$ electromagnetic charge renormalization. It will be
assumed that the external fermion has $t_3=+1/2$ in which case the fermion
to which the photon couples has $t_3=-1/2$. We will denote the charge of
the external fermion by $Q$. After obtaining the tensor integral for the
Feynman diagram by usual methods, applying the projection operator as in
eq.(\ref{eq:vertexproj}) gives the scalar integral representation
\begin{eqnarray*}
 & &
i\gamma_\mu\gamma_L g\left(\frac{g^2}{16\pi^2}\right)^2 Qs_\theta
    \frac{(n-2)}{2n}\int\frac{d^np}{i\pi^2}\int\frac{d^nq}{i\pi^2}\
    \frac{[3p^2q^2-2(p\cdot q)^2+q^2(p\cdot q)]}
         {p^2[p^2+M_W^2]^2 q^4(q+p)^2}\\
 & & \ \ \ \ \
=i\gamma_\mu\gamma_L g\left(\frac{g^2}{16\pi^2}\right)^2 Q s_\theta
     \frac{(n-2)}{2n}\\
& & \ \ \ \ \ \ \ \ \ \times
\left\{-\frac{1}{2}I(1,2,2,-1,n,M_W^2)
-\frac{1}{2}I(-1,2,2,1,n,M_W^2)
+\frac{3}{2}I(0,2,1,1,n,M_W^2)\right\}\\
 & & \ \ \ \ \
=i\gamma_\mu\gamma_L g\left(\frac{g^2}{16\pi^2}\right)^2
   \frac{Q}{2} s_\theta(\pi M_W^2)^{n-4}
   \Gamma(4-n)\Gamma\left(2-\frac{n}{2}\right)\Gamma\left(\frac{n}{2}-1\right)
\end{eqnarray*}
exactly. These diagrams and their related topologies do not vanish when summed
over a complete fermion generation.

\section{Box Diagrams}

For 4-fermion processes one-loop box diagrams are finite but at
${\cal O}(N_f\alpha^2)$ they develop logarithmic divergences.
A useful set of identities for calculating one-loop box diagrams appears
in ref.\cite{SirlinBox}. They are, however, valid only for $n=4$
because of their intended use at one-loop. For general $n$ it may be shown
that these relations become
\begin{eqnarray}
{[}\gamma_\rho\gamma_\mu\gamma_\sigma\gamma_{L,R}{]}_1
{[}\gamma_\rho\gamma_\nu\gamma_\sigma\gamma_{L,R}{]}_2
&=&4\delta_{\mu\nu}{[}\gamma_\alpha\gamma_{L,R}{]}_1
                       {[}\gamma_\alpha\gamma_{L,R}{]}_2
                 +(n-4){[}\gamma_\mu\gamma_{L,R}{]}_1
                       {[}\gamma_\nu\gamma_{L,R}{]}_2\ \ \
\label{eq:firstBoxid}\\
{[}\gamma_\rho\gamma_\mu\gamma_\sigma\gamma_{L,R}{]}_1
{[}\gamma_\rho\gamma_\nu\gamma_\sigma\gamma_{R,L}{]}_2
                   &=&4{[}\gamma_\nu\gamma_{L,R}{]}_1
                       {[}\gamma_\mu\gamma_{R,L}{]}_2
                 +(n-4){[}\gamma_\mu\gamma_{L,R}{]}_1
                       {[}\gamma_\nu\gamma_{R,L}{]}_2\ \ \ \\
{[}\gamma_\rho\gamma_\mu\gamma_\sigma\gamma_{L,R}{]}_1
{[}\gamma_\sigma\gamma_\nu\gamma_\rho\gamma_{L,R}{]}_2
                   &=&4{[}\gamma_\nu\gamma_{L,R}{]}_1
                       {[}\gamma_\mu\gamma_{L,R}{]}_2
                 +(n-4){[}\gamma_\mu\gamma_{L,R}{]}_1
                       {[}\gamma_\nu\gamma_{L,R}{]}_2\ \ \ \\
{[}\gamma_\rho\gamma_\mu\gamma_\sigma\gamma_{L,R}{]}_1
{[}\gamma_\sigma\gamma_\nu\gamma_\rho\gamma_{R,L}{]}_2
&=&4\delta_{\mu\nu}{[}\gamma_\alpha\gamma_{L,R}{]}_1
                       {[}\gamma_\alpha\gamma_{R,L}{]}_2
                 +(n-4){[}\gamma_\mu\gamma_{L,R}{]}_1
                       {[}\gamma_\nu\gamma_{R,L}{]}_2\ \ \ \ \
\label{eq:fourthBoxid}
\end{eqnarray}
where the square brackets $[\ ]_1$ and $[\ ]_2$ indicate that the enclosed
$\gamma$-matrices are associated with the external fermion currents
$J_1$ and $J_2$ respectively.

The general ${\cal O}(N_f\alpha^2)$ box diagram for massless external
fermions at $q=0$ therefore takes the form
\begin{equation}
B_{\mu\nu}J_{1\mu}J_{2\nu}=B\cdot J_{1\alpha}J_{2\alpha}.
\end{equation}
Here $B_{\mu\nu}$ is a tensor integral and the product of
$J_{1\alpha}J_{2\alpha}$ can be constructed from one or a combination of
the $\gamma$-matrices appearing in
eq.(\ref{eq:firstBoxid})--(\ref{eq:fourthBoxid}). As for the self-energy
contributions, the tensor integral $B_{\mu\nu}$, can only be proportional
to $\delta_{\mu\nu}$ and hence, using the projection operator method of
section~2, the scalar integral representation for the form factor $B$ is
seen to be $B=(\delta_{\mu\nu}/n)B_{\mu\nu}=B_{\mu\mu}/n$.
In this case the identities
(\ref{eq:firstBoxid})--(\ref{eq:fourthBoxid}) simplify to become
\begin{eqnarray}
\frac{\delta_{\mu\nu}}{n}
{[}\gamma_\rho\gamma_\mu\gamma_\sigma\gamma_{L,R}{]}_1
{[}\gamma_\rho\gamma_\nu\gamma_\sigma\gamma_{L,R}{]}_2
&=&\frac{(5n-4)}{n}{[}\gamma_\alpha\gamma_{L,R}{]}_1
                   {[}\gamma_\alpha\gamma_{L,R}{]}_2\\
\frac{\delta_{\mu\nu}}{n}
{[}\gamma_\rho\gamma_\mu\gamma_\sigma\gamma_{L,R}{]}_1
{[}\gamma_\rho\gamma_\nu\gamma_\sigma\gamma_{R,L}{]}_2
                   &=&{[}\gamma_\alpha\gamma_{L,R}{]}_1
                       {[}\gamma_\alpha\gamma_{R,L}{]}_2
\label{eq:secondBoxid}\\
\frac{\delta_{\mu\nu}}{n}
{[}\gamma_\rho\gamma_\mu\gamma_\sigma\gamma_{L,R}{]}_1
{[}\gamma_\sigma\gamma_\nu\gamma_\rho\gamma_{L,R}{]}_2
                   &=&{[}\gamma_\alpha\gamma_{L,R}{]}_1
                       {[}\gamma_\alpha\gamma_{L,R}{]}_2\\
\frac{\delta_{\mu\nu}}{n}
{[}\gamma_\rho\gamma_\mu\gamma_\sigma\gamma_{L,R}{]}_1
{[}\gamma_\sigma\gamma_\nu\gamma_\rho\gamma_{R,L}{]}_2
&=&\frac{(5n-4)}{n}{[}\gamma_\alpha\gamma_{L,R}{]}_1
                   {[}\gamma_\alpha\gamma_{R,L}{]}_2.
\end{eqnarray}

\subsection{Example}

The diagram shown in Fig.\ref{fig:example}(b) occurs in the ${\cal O}(N_f\alpha^2)$
corrections to muon decay. Using (1) and
(\ref{eq:secondBoxid}) the resulting the Feynman diagram can be written as
\begin{eqnarray*}
& &
\frac{g^2}{16c_\theta^2}\left(\frac{g^2}{16\pi^2}\right)^2
{[}\gamma_\alpha\gamma_L{]}_\mu{[}\gamma_\alpha\gamma_L{]}_e
\frac{(n-2)}{(n-1)}\int\frac{d^np}{i\pi^2}\int\frac{d^nq}{i\pi^2}
\frac{1}{[p^2+M_Z^2][p^2+M_W^2]^2 q^2(q+p)^2}\\
 & & \ \ \ \ \ \ \ \ \
=\frac{g^2}{16c_\theta^2}\left(\frac{g^2}{16\pi^2}\right)^2
{[}\gamma_\alpha\gamma_L{]}_\mu{[}\gamma_\alpha\gamma_L{]}_e
\frac{(n-2)}{(n-1)}\\
 & & \ \ \ \ \ \ \ \ \ \ \ \ \ \ \ \times
\frac{1}{(M_Z^2-M_W^2)^2}\Big\{I(0,1,1,1,n,M_Z^2)-I(0,1,1,1,n,M_W^2)\\
 & & \ \ \ \ \ \ \ \ \ \ \ \ \ \ \ \ \ \ \ \ \ \ \ \ \ \ \ \ \ \ \ \ \ \ \
                +(M_Z^2-M_W^2)I(0,2,1,1,n,M_W^2)
\Big\}\\
 & & \ \ \ \ \ \ \ \ \
=-\frac{g^2}{24 M_W^2s_\theta^4}
\left(\frac{g^2}{16\pi^2}\right)
{[}\gamma_\alpha\gamma_L{]}_\mu{[}\gamma_\alpha\gamma_L{]}_e
(\pi M_W^2)^{-2\epsilon}\Gamma(\epsilon)\\
 & & \ \ \ \ \ \ \ \ \ \ \ \ \ \ \ \times
\left\{(s_\theta^2+\ln c_\theta^2)
     \left(1+\epsilon\left(\frac{8}{3}-\gamma\right)\right)
            +\epsilon\ln^2 c_\theta^2\right\}
\end{eqnarray*}
up to ${\cal O}(1)$ in $\epsilon=2-n/2$.

\section{Derivatives of Self-energy Diagrams}

The derivatives of self-energies with respect to the square of the
external momentum, $k^2$, are required for the calculation physical
processes. They are often thought of as wavefunction renormalization
factors although they occur even when wavefunction renormalization
has not been explicitly performed \cite{ZMass2}.
As we are concerned with low-energy processes and massless fermions,
these derivatives will only be required at $k^2=0$.
In that case the scalar integral representation may be obtained by the
following method.

For a vector boson the self-energy may be written in the form
\begin{equation}
\Pi_{\mu\nu}(k^2)=\left(\delta_{\mu\nu}-\frac{k_\mu k_\nu}{k^2}\right)\Pi_T(k^2)
                 +\left(\frac{k_\mu k_\nu}{k^2}\right)\Pi_L(k^2)
\end{equation}
with $\Pi_T$ and $\Pi_L$ being the transverse and longitudinal form-factors
respectively.
\begin{eqnarray}
\Pi_T(k^2)&=&\frac{1}{n-1}\left(\delta_{\mu\nu}-\frac{k_\mu k_\nu}{k^2}\right)
             \Pi_{\mu\nu}(k^2)\\
\Pi_L(k^2)&=&\left(\frac{k_\mu k_\nu}{k^2}\right)\Pi_{\mu\nu}(k^2).
\end{eqnarray}

For massless fermions self-energy diagrams take the form
\begin{equation}
\Sigma(k)=i\slash{k}\left(F_L(k^2)\gamma_L+F_R(k^2)\gamma_R\right)
\end{equation}
and
\begin{equation}
F_{L,R}(k^2)=-\frac{i}{2k^2}\Tr\left\{\Sigma(k)\slash{k}\gamma_{R,L}\right\}
\end{equation}

In either case given expressions for the self-energies in the form of tensor
integrals, the scalar integral representations for the corresponding
form factors $\Pi_L$, $\Pi_T$ or $F_L$, $F_R$ are easily obtained.
In general, then, these scalar integral representations may be written
\begin{equation}
F(k^2)=\int\frac{d^np}{i\pi^2}\int\frac{d^nq}{i\pi^2}
      \frac{F(k^2,p^2,q^2,k\cdot p,k\cdot q,p\cdot q)}
           {[(k+p)^2+M_1^2]^{N_1}[(k+q)^2+M_2^2]^{N_2}[(k+p+q)^2+M_3^2]^{N_3}}
\ \ \ \ \ \ \ {}
\label{eq:ScalarIntRep}
\end{equation}
where $N_1$, $N_2$ and $N_3$ are integer powers.

In the region of small $k^2$ we may expand the propagators as a power series
in $(k\cdot p)$ and $(k\cdot q)$. Thus, for example,
\begin{eqnarray*}
\frac{1}{[(k+p)^2+M_1^2]^{N_2}}&=&\frac{1}{[k^2+p^2+M_1^2]^{N_2}}
                          \left(1-2N_2\frac{(k\cdot p)}{k^2+p^2+M_1^2}
                                 +...\right)
\label{eq:expansion}\\
\frac{1}{[(k+p+q)^2+M_3^2]^{N_3}}&=&\frac{1}{[k^2+(p+q)^2+M_3^2]^{N_3}}
                      \left(1-2N_3\frac{(k\cdot p)+(k\cdot q)}
                                   {k^2+(p+q)^2+M_3^2}
                             +...\right)\\
\end{eqnarray*}
with the result that the integral in eq.(\ref{eq:ScalarIntRep}) becomes
\begin{equation}
F(k^2)=\int\frac{d^np}{i\pi^2}\int\frac{d^nq}{i\pi^2}
       \sum_{i,j}(k\cdot p)^i (k\cdot q)^j F_{ij}(k^2,q^2,p^2).
\end{equation}
Upon integration terms in which $i+j$ is odd vanish. Terms in the integrand
with $i+j=2$ are $(k\cdot v_1)(k\cdot v_2)F(k^2,q^2,p^2)$ in which
$v_1$ and $v_2$ are either $p$ or $q$. Because $F$ is a function only
of squared momenta the integration of $p_\mu q_\nu F(k^2,q^2,p^2)$ can only
yield
a result proportional to $\delta_{\mu\nu}$. Its coefficient may be extracted
using the projection operator $\delta_{\mu\nu}/n$ as in section 2. It follows
then that the terms $(k\cdot v_1)(k\cdot v_2)F(k^2,q^2,p^2)$ in the
integrand may be replaced by $k^2(v_1\cdot v_2)/n\,F(k^2,q^2,p^2)$ without
changing the integral.

In a similar way for $i+j=4$ one can make the replacement
\begin{align*}
(k\cdot v_1)(k\cdot v_2)&(k\cdot v_3)(k\cdot v_4)F(k^2,q^2,p^2)
\longrightarrow\\
&k^4\frac{(v_1\cdot v_2)(v_3\cdot v_4)
         +(v_1\cdot v_3)(v_2\cdot v_4)
         +(v_1\cdot v_4)(v_2\cdot v_3)}{n(n+2)}F(k^2,q^2,p^2)
\end{align*}
where again the $v_i$ are $p$ or $q$.

The integral is thus transformed into a form whose only $k$ dependence
is enters only through $k^2$.

Differentiation of the integrand with respect to $k^2$ can now be
performed and setting $k^2=0$ yields a result expressible in terms of the
master integral (\ref{eq:MasterIntexpr}).

For most practical purposes $i+j=4$ is as high as one needs to go in the
series expansion. The general term in this series can be found in
\cite{DavyTausk} and requires repeated application of the momentum-space
d'Alembertian operator.

\subsection{Example}

A self-energy diagram for a massless fermion is shown in Fig.1c and
may be written
\[
\Sigma(k)=i\slash{k}A_L(k^2)\gamma_L.
\]
After obtaining the scalar integral representation for $A_L(k^2)$ and
expanding the propagator as in (\ref{eq:expansion}) up to $(k\cdot p)^2$
one obtains
\begin{eqnarray*}
A_L(k^2)&=&-\frac{1}{4k^2}\left(\frac{g^2}{16\pi^2}\right)^2
\frac{(n-2)}{(n-1)}
\int\frac{d^np}{i\pi^2}\int\frac{d^nq}{i\pi^2}
\frac{1}{[p^2+M_W^2]^2 q^2(q+p)^2}\\
& &\times
\left\{\frac{(3-n)k^2}{k^2+p^2}
      +\frac{2(p^2-k^2)(k^2+(n-2)p^2)}{p^2(k^2+p^2)^3}(k\cdot p)^2
\right\}
\end{eqnarray*}
After the replacement $(k\cdot p)^2\rightarrow k^2p^2/n$ the derivative
with respect to $k^2$ at $k^2=0$ is found to be
\begin{eqnarray*}
\frac{\partial A_L(k^2)}{\partial k^2}\Bigg\vert_{k^2=0}
&=&\left(\frac{g^2}{16\pi^2}\right)^2\frac{(n-2)(n-4)}{4n}
\int\frac{d^np}{i\pi^2}\int\frac{d^nq}{i\pi^2}
\frac{1}{[p^2+M_W^2]^2 q^2 (q+p)^2}\ \ \ \ \\
&=&\left(\frac{g^2}{16\pi^2}\right)^2\frac{(n-2)(n-4)}{4n} I(0,2,1,1,n,M_W^2)\\
&=&-\left(\frac{g^2}{16\pi^2}\right)^2(\pi M_W^2)^{n-4}\frac{1}{2n}\Gamma(5-n)
                \Gamma\left(2-\frac{\textstyle n}{\textstyle 2}\right)
                \Gamma\left(\frac{\textstyle n}{\textstyle 2}-1\right)
\end{eqnarray*}
exactly. This result is related to vertex diagrams by Ward identities that
serve as a useful check.

\section{Acknowledgments}
This work was supported in part by the U.S. Department of Energy.

\end{document}